\documentclass[letterpaper,english,preprint,aps,nofootinbib]{revtex4-1}
\usepackage[T1]{fontenc}
\usepackage[latin9]{inputenc}
\usepackage{color}
\usepackage{amsmath}
\usepackage{amssymb}
\usepackage{stackrel}
\usepackage{graphicx}
\usepackage{float}
\usepackage{subfigure}
\usepackage{xcolor}
\makeatletter
\@ifundefined{textcolor}{}
{%
 \definecolor{BLACK}{gray}{0}
 \definecolor{WHITE}{gray}{1}
 \definecolor{RED}{rgb}{1,0,0}
 \definecolor{GREE}{rgb}{0,1,0}
 \definecolor{BLUE}{rgb}{0,0,1}
 \definecolor{CYAN}{cmyk}{1,0,0,0}
 \definecolor{MAGENTA}{cmyk}{0,1,0,0}
 \definecolor{YELLOW}{cmyk}{0,0,1,0}
}

\makeatother

\usepackage{babel}
\begin{document}
\global\long\def\a{\alpha}%
\global\long\def\b{\beta}%
\global\long\def\c{\chi}%
\global\long\def\d{\delta}%
\global\long\def\e{\epsilon}%
\global\long\def\f{\phi}%
\global\long\def\g{\gamma}%
\global\long\def\h{\eta}%
\global\long\def\i{\iota}%
\global\long\def\k{\kappa}%
\global\long\def\l{\lambda}%
\global\long\def\m{\mu}%
\global\long\def\n{\nu}%
\global\long\def\o{\omega}%
\global\long\def\p{\pi}%
\global\long\def\q{\theta}%
\global\long\def\r{\rho}%
\global\long\def\s{\sigma}%
\global\long\def\t{\tau}%
\global\long\def\u{\upsilon}%
\global\long\def\x{\xi}%
\global\long\def\y{\psi}%
\global\long\def\z{\zeta}%

\global\long\def\ve{\varepsilon}%
\global\long\def\vf{\varphi}%
\global\long\def\vs{\varsigma}%
\global\long\def\vq{\vartheta}%

\global\long\def\D{\Delta}%
\global\long\def\F{\Phi}%
\global\long\def\G{\Gamma}%
\global\long\def\L{\Lambda}%
\global\long\def\Q{\Theta}%
\global\long\def\S{\Sigma}%
\global\long\def\U{\Upsilon}%
\global\long\def\W{\Omega}%
\global\long\def\X{\Xi}%
\global\long\def\Y{\Psi}%

\global\long\def\6{\partial}%
\global\long\def\8{\infty}%
\global\long\def\j{\int}%
\global\long\def\w{}%
\global\long\def\R{\Rightarrow}%
\global\long\def\*{\times}%
\global\long\def\={\equiv}%
\global\long\def\.{\cdot}%

\global\long\def\cA{\mathcal{A}}%
\global\long\def\cD{\mathcal{D}}%
\global\long\def\cF{\mathscr{\mathcal{F}}}%
\global\long\def\cH{\mathcal{H}}%
\global\long\def\cL{\mathcal{L}}%
\global\long\def\cJ{\mathcal{J}}%
\global\long\def\cO{\mathcal{O}}%
\global\long\def\cP{\mathcal{P}}%
\global\long\def\cQ{\mathcal{Q}}%
\global\long\def\cY{\mathcal{Y}}%

\global\long\def\sA{\mathscr{A}}%
\global\long\def\sD{\mathscr{D}}%
\global\long\def\sF{\mathscr{F}}%
\global\long\def\sH{\mathscr{H}}%
\global\long\def\sL{\mathscr{L}}%
\global\long\def\sJ{\mathscr{J}}%
\global\long\def\sO{\mathscr{O}}%
\global\long\def\sP{\mathscr{P}}%
\global\long\def\sQ{\mathscr{Q}}%
\global\long\def\sY{\mathscr{Y}}%

\global\long\def\na{\nabla}%
\global\long\def\cd{\cdots}%
\global\long\def\da{\dagger}%
\global\long\def\ot{\otimes}%
\global\long\def\we{\wedge}%
\global\long\def\qu{\quad}%

\global\long\def\fL{\mathfrak{L}}%
\global\long\def\md{\mathrm{d}}%
\global\long\def\re{\mathrm{Re}}%
\global\long\def\im{\mathrm{Im}}%
\global\long\def\hb{\hbar}%

\title{Phase transition of modified thermodynamics of the Kerr-AdS black hole}






\author{Qin Liu$^{1}$}\email{qliu@st.gxu.edu.cn}
\author{Xiaoning Wu$^{3,4}$}\email{wuxn@amss.ac.cn}
\author{Xiao Zhang$^{2,3,4}$}\email{xzhang@amss.ac.cn}

\affiliation{$^1$School of Physical Science and Technology, Guangxi University, Nanning 530004, Guangxi, People's Republic of China}
\affiliation{$^2$Guangxi Center for Mathematical Research, Guangxi University, Nanning 530004, Guangxi, People's Republic of China }
\affiliation{$^3$Institute of Mathematics, Academy of Mathematics and Systems Science and Hua Loo-Keng Laboratory, Chinese Academy of Sciences, Beijing 100190, People's Republic of China}
\affiliation{$^4$School of Mathematical Sciences, University of Chinese Academy of Sciences, Beijing 100049, People's Republic of China}

\begin{abstract}
	We investigate the critical phenomena of Kerr-AdS black holes in the modified first law of thermodynamics. The modified black hole thermodynamics exhibits the van der Waals-like phase structure. All the critical exponents are calculated, and the swallowtail diagram of free energy is plotted. Comparing with existing results, the main difference is the correspondence between thermal quantities of the Kerr-AdS black holes and the van der Waals system. In previous work\cite{TWY}, the correspondence was $(\Omega_{H},J)$$\rightarrow(V,P)$. In our work, the correspondence is $(J,\hat{\Omega}_{H})$$\rightarrow(V,P)$. This difference results from  rotating effect. The modified black hole thermodynamics is associated with rotating observers. The free energy in such reference contains an extra rotating energy. This extra part induces a Legendre transformation in $(\hat{\Omega}_{H},J)$ cross-section, which yields the different correspondence.
	
\end{abstract}
\maketitle

\section{introduction}
Since Bekenstein and Hawking \cite{BE,H} found that black holes can be seen as thermodynamic objects with temperature, it is widely accepted that black hole thermodynamics has been an exciting and challenging field in theoretical physics. As fascinating phenomena in black hole thermodynamics, phase transitions and critical behaviors have attracted the attention of many researchers. Based on the analogy between black holes and thermodynamic objects, Hawking and Page proved the existence of a phase transition between the Schwarzschild-AdS black hole and the thermal AdS space \cite{HP}. After their pioneering research, many works were carried out in this direction and rich phase structures were discovered \cite{AL,CCS,CLZ,CV,L}. The later-established AdS/CFT correspondence has further stimulated attention to the asymptotically anti-de Sitter (AdS) black holes \cite{W1,CEJM}. In different AdS black hole backgrounds, phase structures and critical phenomena have been studied and promoted \cite{BMS1,CCK,N,W2,SSS,TWY,WL,CWL,ML}.

It is worth noting that the energy definition of Kerr-AdS black holes is not unique. This is because the standard Komar energy expression diverges at infinity. Henneaux and Teitelboim \cite{HT} first derived the energy $m/\Xi^{2}$, which satisfies the first law of black hole thermodynamics. The authors \cite{CJK} proposed the energy $m/\Xi^{3/2}$ by applying the Hamiltonian approach. Using the Iyer-Wald formalism, Gao et al. \cite{GDG} clarified the origins of the two different energy definitions of Kerr-AdS black holes. They found the energies associated with different observers at infinity,  and there is a relative rotation between the two kinds of observers.  The associated thermodynamics for the different energies are also obtained in \cite{GDG}. For non-rotating observers, the energy is $m/\Xi^{2}$, and for rotating observers, the energy is $m/\Xi^{3/2}$. The energy $m/\Xi^{2}$ corresponds to the standard thermodynamics of black hole, and its related critical phenomena have been studied, including phase transition and critical exponents \cite{TWY}. The other energy $m/\Xi^{3/2}$ is related to the modified first law of  black hole thermodynamics \cite{GDG}. Given that the phase transition is associated with the degrees of freedom within the system, and there is only a relative rotation between two observers, it is reasonably conjectured that there should be similar phase structures in the modified thermodynamics. In the current work, we try to investigate whether the phase transition structure still exists under this modified first law of thermodynamics. Furthermore, the observer dependence of the phase structure will be discussed.

The paper is organized as follows. Section II briefly reviews the basic knowledge of the Kerr-AdS black hole. Section III is mainly concerned with the modified first law and thermodynamic quantities. The main results are in section IV and V, the phase structure of the modified first law has been exhibited. Discussions are given in the last section.

\section{Kerr-AdS black hole thermodynamics}

The Kerr-AdS black hole solution of the Einstein equations in the Boyer-Lindquist coordinates reads
\begin{equation}
	\begin{aligned}
		\mathrm{d}s^{2}=-&\frac{\Delta}{\rho^2}\big( \mathrm{d}t-\frac{a\sin^2\theta}{\Xi}\mathrm{d}\phi \big)^2+\frac{\rho^2}{\Delta}\mathrm{d}r^{2}\\+&\frac{\rho^2}{\Sigma}\mathrm{d}\theta^{2}+\frac{\Xi\sin^2\theta}{\rho^2}\big( a\mathrm{d}t-\frac{r^2+a^2}{\Xi}\mathrm{d}\phi \big)^2,
	\end{aligned}
\end{equation}

with
\begin{equation}
	\begin{aligned}
		\rho^2=&r^2+a^2\cos^2\theta,\\
		\Xi=&1-\frac{a^2}{l^2},\\
		\Sigma=&1-\frac{a^2}{l^2}\cos^2\theta,\\
		\Delta=&(r^2+a^2)(1+\frac{r^2}{l^2})-2mr.\\
	\end{aligned}
\end{equation}
Here $\Lambda=-3l^{-2}$ is the cosmological constant, $m, a$ are the mass parameter and angular momentum parameter. The associated thermodynamic quantities are \cite{TWY}:
\begin{align}
	T=\frac{3r_{+}^4+(a^2+l^2)r_{+}^2-l^2a^2}{4\pi l^2r_{+}(r_{+}^2+a^2)},
	\quad S=\frac{\pi(r_{+}^2+a^2)}{\Xi},
	\quad \Omega_{H}=\frac{a\Xi}{r_{+}^2+a^2},
	\label{2.3}
\end{align}
where $r_{+}$ represents the horizon radius satisfying $\Delta(r_{+})=0$, $T$, $S$ and $\Omega_{H}$ are defined as the Hawking temperature, the Bekenstein-Hawking entropy and the angular velocity respectively. The energy $M$ and the angular momentum $J$ are $M=\frac{m}{\Xi^2}$ and $J=\frac{am}{\Xi^2}$ respectively.

In 2012, S. Gunasekaran, D. Kubiz$\check{n}$�k, and R. B.Mann introduced the extended phase space for AdS black hole \cite{GKM}, where the cosmological constant  $\Lambda$ can be interpreted as a pressure term via the following relation 
\begin{align}
	P=-\frac{\Lambda}{8\pi}.
\end{align}
The first law of the black hole thermodynamics and the Smarr formula are
\begin{align}
	\delta M=&T\delta S+\Omega_{H}\delta J+V\delta P,\\
	\frac{M}{2}=&TS+\Omega_{H}J-VP,
\end{align}
where the thermodynamic volume V is $\frac{4\pi l^2 r_{+}(a^2+r_{+}^2)}{3(-a^2+l^2)}$. Followed Gunasekaran et al.'s perspective, many works had been down to explore the thermodynamic properties of AdS black holes on the extended phase space \cite{ZZMZ,WL1,XXZ,LS}.

\section{modified black hole thermodynamics}

It is known that there is ambiguity on energy notion of the Kerr-AdS black holes \cite{HT,CJK}. This ambiguity has been a long standing issue. In the very recent article \cite{GDG}, the authors proposed a natural criteria to justify the notion of energy. In particular, they examined whether the associated first law of the black hole thermodynamics exists. Within the Iyer-Wald formalism, two versions of the first law and the Smarr formula were established for different energies. The difference originated from the choice of the Killing vectors. The standard energy notion $m/\Xi^{2}$ is associated with the Killing vector $\frac{\partial}{\partial T}=\frac{\partial}{\partial t}+\frac{1}{3}a\Lambda \frac{\partial}{\partial \phi}$. The relevant thermodynamic quantities, the first law  and the standard results are presented in section II. The other energy notion $\hat{M}=m/\Xi^{3/2}$ is related to the Killing vector $\frac{1}{\sqrt{\Xi}}\frac{\partial}{\partial t}$  , using the notation from \cite{GDG}, the corresponding thermodynamic quantities are
\begin{align}
	\hat{\Omega}_{H}=\frac{\Omega_{H}}{\sqrt{\Xi}},
	\quad \hat{T}=\frac{T}{\sqrt{\Xi}},
	\quad \hat{V}=\frac{V}{\sqrt{\Xi}}.
	\label {3.1}
\end{align}
The modified first law of black hole thermodynamics is presented as
\begin{align}
	\delta\hat{M}=\hat{T}\delta S+	\hat{\Omega}_{H}\delta J+\hat{V}\delta P.	
\end{align}
Based on the concrete expressions of the Killing vectors, there exists a relative rotation between two kinds of observers. Unlike a naive observation, $\frac{\partial}{\partial T}$ coincides with the generator of the conformal boundary, corresponding to non-rotating observers, whereas rotating observers are related to $\frac{\partial}{\partial t}$.

\section{phase structure on ($\hat{\Omega}_{H}$, $J$) section}

Using eq.(\ref{3.1}), the functions $\hat{\Omega}_{H}$,$J$ and $\hat{T}$ can be expressed as
\begin{align}
	J=&\frac{3a(a^2+r_{+}^2)(3+8P\pi r_{+}^2 )}{2(3-8a^2P \pi)^2r_{+}},\label{4.1}\\
	\hat{\Omega}_{H}=&\frac{2a\sqrt{P(-a^2+\frac{3}{8P\pi})}\sqrt{\frac{2\pi}{3}}}{a^2+r_{+}^2},\label{4.2}\\
	\hat{T}=&\frac{\sqrt{\frac{1}{9-24a^2P\pi}}(a^2(-3+8P\pi r_{+}^2)+3(r_{+}^2+8P\pi r_{+}^4))}{4\pi r_{+}(a^2+r_{+}^2) }.\label{4.3}		
\end{align}
The above equations can be viewed as equation of state of the Kerr-AdS black hole.

In this section, we focus on the phase structure of the modified black hole thermodynamics on the $(\hat{\Omega}_{H},J)$ plane. Previous work \cite{TWY} had considered similar problems for the standard black hole thermodynamics on the $(J,\Omega_{H})$ section, without considering the extended phase space.

\begin{figure}[H]
	\centering
	\includegraphics[width=0.7 \textwidth]{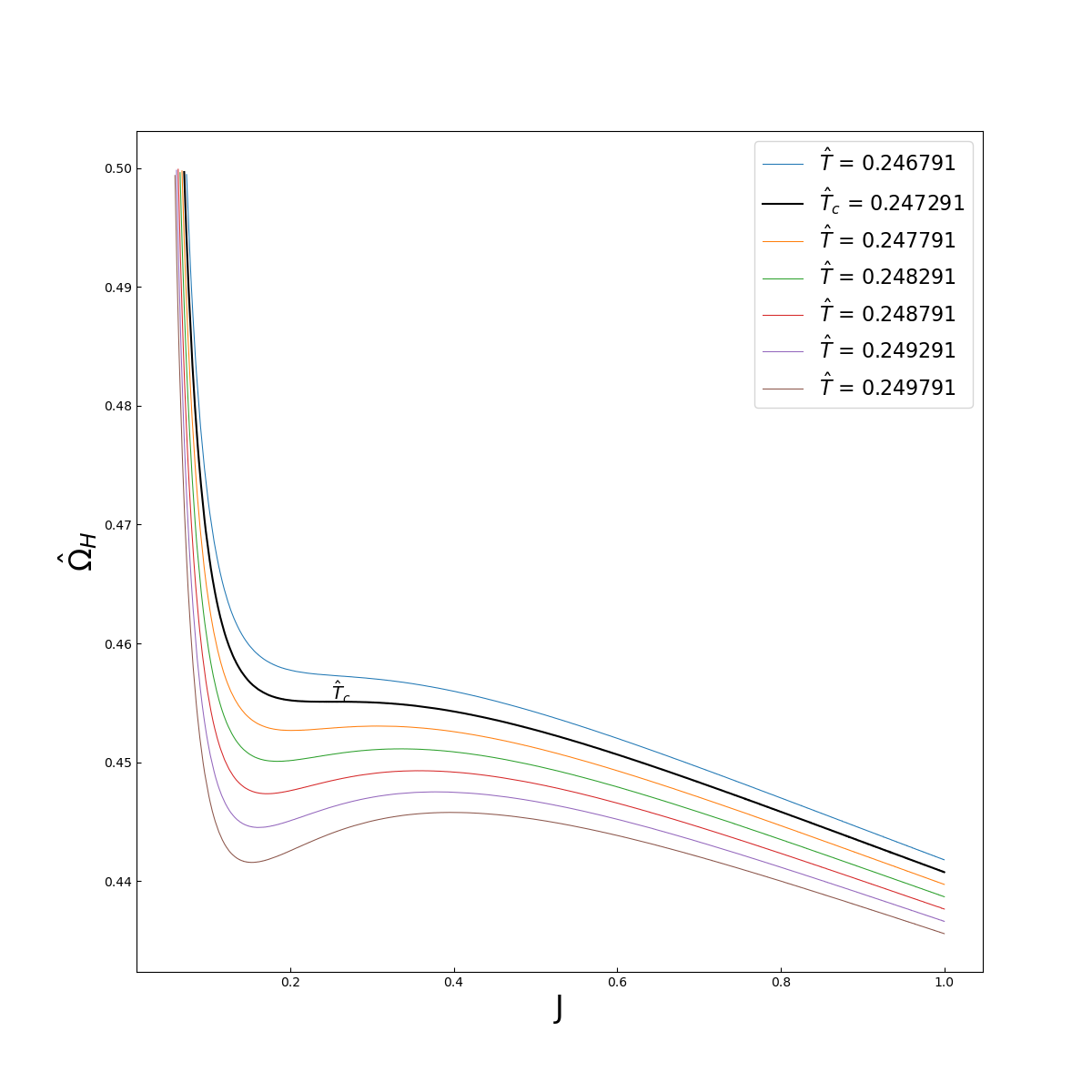}
	\caption{We draw isotherms of Kerr-AdS black hole with $l=1$ in $\hat{\Omega}_{H}-J$ plane. Qualitatively, the isotherms near the critical temperature $\hat{T}_{c}$ have similar behavior of van der Waals system. The black solid line is the critical isotherm.}.
	\label{Fig.1}
\end{figure}

According to the  equation of state (\ref{4.1})-(\ref{4.3}), the isotherms for various temperatures on $\hat{\Omega}_{H}-J$ plane are presented in Fig.\ref{Fig.1}.
Qualitatively, Fig.\ref{Fig.1} is similar to the liquid/gas PVT diagram \cite{R}. Under the following correspondence,
\begin{eqnarray}
	\hat{\Omega}_{H}\rightarrow P, J\rightarrow V,\label{4.4}
\end{eqnarray}
there is a van der Waals-like phase structure. By contrast, the correspondence in the previous article \cite{TWY} was $J\rightarrow P,  \Omega_{H}\rightarrow V$.

As the temperature decreases until $\hat{T}_{c}$, an inflection point is formed. So, the temperature $\hat{T}_{c}$ and the angular velocity $\hat{\Omega}_{H}$ at the critical point satisfy  \cite{R}
\begin{align}
	\left(\frac{\partial \hat{\Omega}_{H} }{\partial J}\right)_{\hat{T}_{c}}=0,
	\quad\left(\frac{\partial^2 \hat{\Omega}_{H} }{\partial J^2}\right)_{\hat{T}_{c}}=0.
\end{align}

Combining with the equation of state (\ref{4.1})-(\ref{4.3}), the critical point is
\begin{align}
	J_{c}=&\frac{3(4+\sqrt{2})(1+2\sqrt{2})^{3/2}\sqrt{-5+4\sqrt{2}}(-6+5\sqrt{2})}{448(-3+\sqrt{2})^2P\pi},\\
	\hat{\Omega}_{Hc}=&\frac{2\sqrt{\frac{2}{3}(-23+17\sqrt{2})\pi P}}{(-2+3\sqrt{2})},\\
	\hat T_{c}=&\frac{7(-2+3\sqrt{2})\sqrt{\frac{2P}{9\pi -3\sqrt{2}\pi}}}{(1+2\sqrt{2})^{3/2}(-6+5\sqrt{2})}.
\end{align}

\subsection{Law of equal area and order parameter}

\begin{figure}[H]
	\centering
	\subfigure[
	]{
		\label{Fig.2}
		\includegraphics[width=0.45\textwidth]{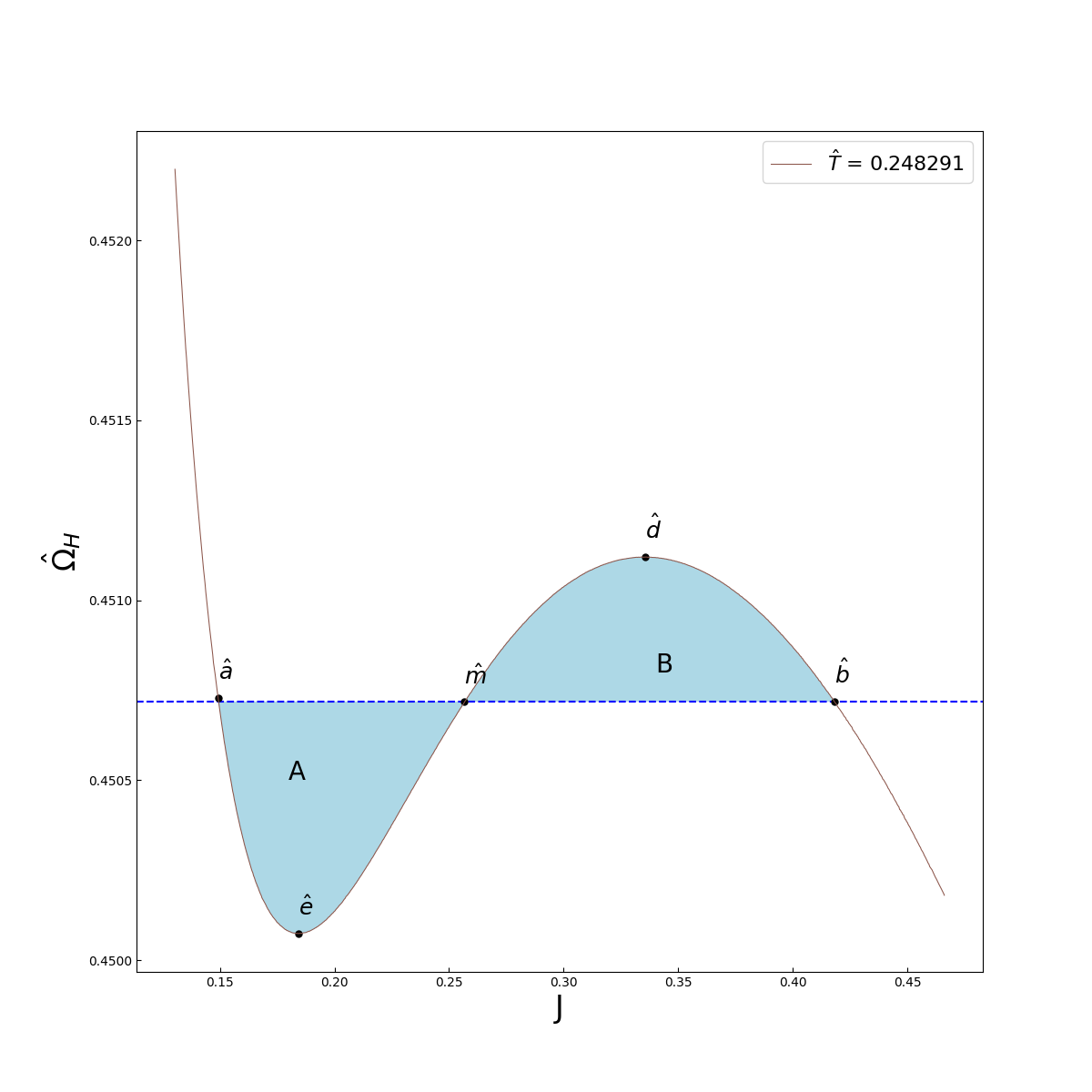}}
	\subfigure[]{
		\label{Fig.3}
		\includegraphics[width=0.45\textwidth]{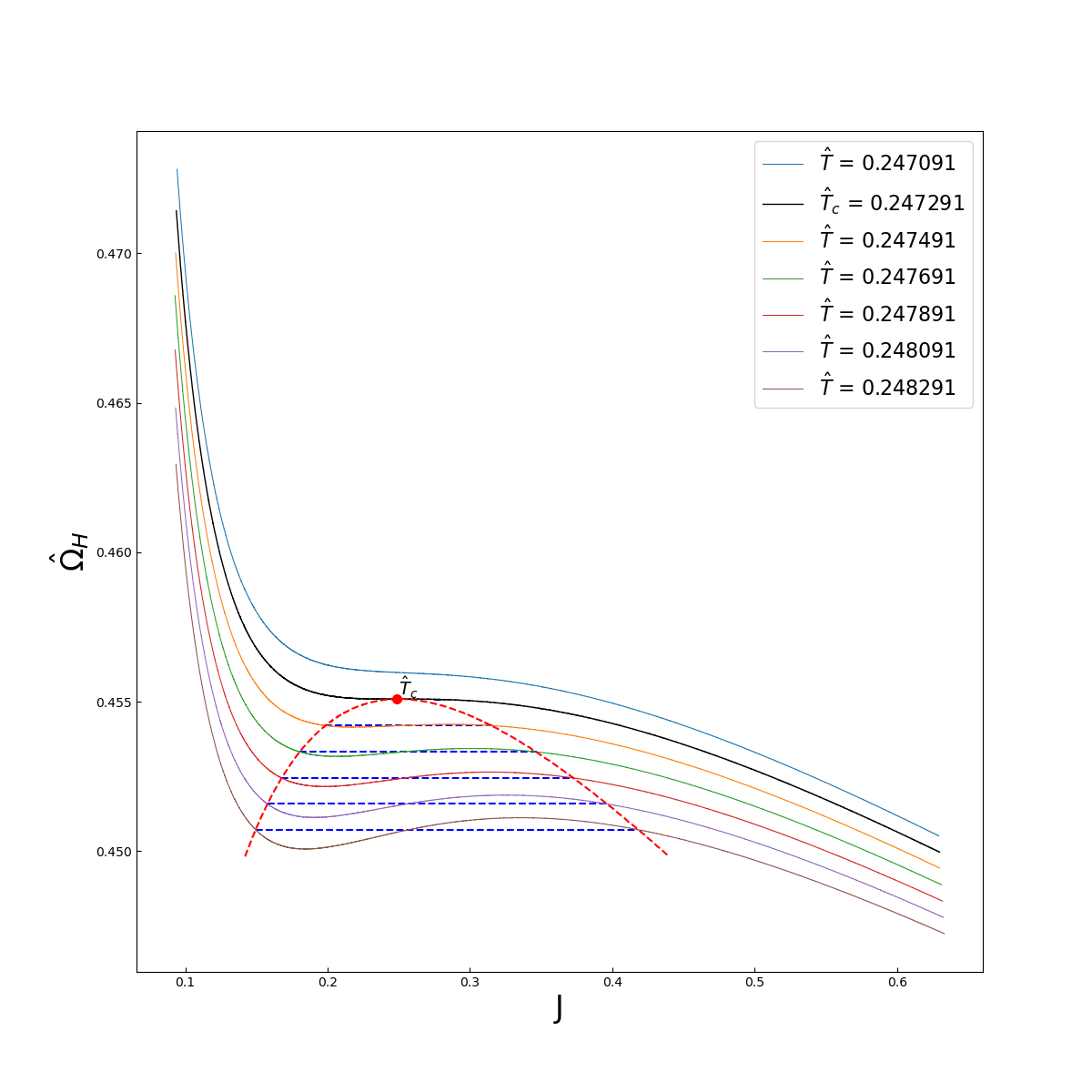}}
	\caption{(a) The brown solid line represents a isotherm with temperature $\hat{T}=0.248291$ and $l=1$, $\hat{a}$ and $\hat{b}$ are coexistence states. Because of equal free energy requirement of coexistence states $G(\hat{a})=G(\hat{b})$, area($A$)=area($B$). (b) The red dashed line is the coexistence curve of phase transition above $\hat{T}_{c}$. The black solid line represents the critical isotherm with $l=1$.}
	\label{Fig.main}
\end{figure}

In this subsection, we study firstly  the coexistence condition between two states, from which the law of equal area will be obtained. For two stable states $\hat{a}$ and $\hat{b}$, it is well-known  that coexistence condition  is the equal free energy \cite{R}, i.e. $G(\hat{a})=G(\hat{b})$. Compared with the PTV system, the free energy for the black hole system should be
\begin{eqnarray}
{\rm d}G=-S{\rm d}\hat{T}+J{\rm d}\hat{\Omega}_{H}.
\end{eqnarray} 
Fixing the temperature, one can get the free energy by integrating above equation along the isothermal curve. The difference of free energy between two states should be $\int_{\hat{a}}^{\hat{b}}J {\rm d}\hat{\Omega}_{H}$(this integral is along the isothermal curve which connects the two states). From Fig.\ref{Fig.2}, since $\hat{a}$ and $\hat{b}$ are coexistent states, it means
\begin{eqnarray}
G(\hat{a})-G(\hat{b})=\int_{\hat{\Omega}_{H\hat{a}}}^{\hat{\Omega}_{H\hat{b}}}J {\rm d}\hat{\Omega}_{H}=0,
\end{eqnarray}
which implies the areas of region A and region B in Fig.\ref{Fig.2} are equal, this is just the Maxwell equal area law.

In order to   investigate the critical behavior near the critical point, we need to define the order parameter. Analogous to the van der Waals system,  $\eta=\frac{J_{b}-J_{a}}{2}$ is defined as the order parameter. The coexistence curve is the red dashed line in Fig.\ref{Fig.3}.

\subsection{Critical exponents}

For van der Waals system, near the critical point, one can get the critical exponents  \cite{R}.  For the correspondence (\ref{4.4}),  the analogous critical exponents can be obtained as following.

$\bullet\ $ Degree of critical isotherm:
\begin{align}\label{cri1}
	\hat{\Omega}_{H}-\hat{\Omega}_{Hc}=A_{\delta}\lvert J-J_{c}\rvert^\delta sign(J-J_{c}),\qquad \hat{T}=\hat{T}_{c}.
\end{align}

$\bullet\ $ Degree of coexistence curve:
\begin{align}
	\eta=-A_{\beta}(\hat{T}-\hat{T}_{c})^\beta,\qquad \hat{T}>\hat{T}_{c}.
\end{align}

$\bullet\ $ Degree of heat capacity($J=J_{c}$):
\begin{equation}
	C_{J}=\left\{
	\begin{aligned}
		A_{\alpha'}\left\{-(\hat{T}-\hat{T}_{c})\right\}^{-\alpha'}\quad \hat{T}<\hat{T}_{c}\\
		A_{\alpha}\left\{+(\hat{T}-\hat{T}_{c})\right\}^{-\alpha} \quad \hat{T}>\hat{T}_{c}.\label{725d3}
	\end{aligned}
	\right
	.
\end{equation}

$\bullet\ $ Degree of isothermal compressibility:
\begin{equation}\label{cri4}
	\kappa{_{T}}=-\frac{1}{J}\left(\frac{\partial J }{\partial \hat{\Omega}_{H}}\right)_{\hat T}=\left\{
	\begin{aligned}
		A_{\gamma'}\left\{-(\hat{T}-\hat{T}_{c})\right\}^{-\gamma'}\quad \hat{T}<\hat{T}_{c}\\
		A_{\gamma}\left\{+(\hat{T}-\hat{T}_{c})\right\}^{-\gamma} \quad \hat{T}>\hat{T}_{c}.
	\end{aligned}
	\right
	.
\end{equation}

In the following of this subseciton, we calculate explicitly the critical exponents for the modified black hole thermodynamics.

\subsubsection{Degree of critical isotherm}

At  critical point, the first and second derivatives of $\hat{\Omega}_{H}$ with respect to $J$ satisfy
\begin{align}
	\big(\frac{\partial \hat{\Omega}_{H} }{\partial J}\big)_{\hat T_{c}}=0,
	\quad\big(\frac{\partial^2 \hat{\Omega}_{H} }{\partial J^2}\big)_{\hat T_{c}}=0.
\end{align}

The third derivative can be calculated as
\begin{align}
	\big(\frac{\partial^3 \hat{\Omega}_{H} }{\partial J^3}\big)_{\hat T_{c}}=-\frac{16384 (\pi P) ^{7/2}}{9 \sqrt{87 \sqrt{2}+123} }\neq 0,
\end{align}
hence $\delta =3$ by the definition of $\delta$ in eq.(\ref{cri1}).

\subsubsection{Degree of coexistence curve}

In Fig.\ref{Fig.3}, we plot the curve of the coexisting states using Maxwell's equal-area law. Along this curve, all the thermodynamical quantities only depend on the temperature. To get the value of the degree of coexistence curve, $\beta$, we need to obtain the relations between $\eta$ and $\hat T$ near the critical point. For this purpose,
we expand $\hat{\Omega}_{H}$ in terms of J and $\hat{T}$ to the third order as
\begin{equation}\label{0722-25}
	\begin{split}
		\hat{\Omega}_{H}-\hat{\Omega}_{Hc}\approx&(\partial_{\hat{T}}\hat{\Omega}_{H})_{J}\vert_{c}(\hat{T}-\hat{T}_{c})+\frac{1}{2}(\partial^2_{\hat{T}}\hat{\Omega}_{H})_{J}\vert_{c}(\hat{T}-\hat{T}_{c})^2\\+&\big(\partial_ {\hat{T}}(\partial _{J}\hat{\Omega}_{H})_{\hat{T}}\big)_{J}\vert_{c}(\hat{T}-\hat{T}_{c})(J-J_{c})+\frac{1}{6}(\partial^3_{\hat{T}}\hat{\Omega}_{H})_{J}\vert_{c}(\hat{T}-\hat{T}_{c})^3\\+&\frac{1}{6}(\partial^3_{J}\hat{\Omega}_{H})_{\hat{T}}\vert_{c}(J-J_{c})^3+\frac{1}{2}\big(\partial^2_ {\hat{T}}(\partial _{J}\hat{\Omega}_{H})_{\hat{T}}\big)_{J}\vert_{c}(\hat{T}-\hat{T}_{c})^2(J-J_{c})\\+&\frac{1}{2}\big(\partial_ {\hat{T}}(\partial^2 _{J}\hat{\Omega}_{H})_{\hat{T}}\big)_{J}\vert_{c}(\hat{T}-\hat{T}_{c})(J-J_{c})^2.
	\end{split}
\end{equation}

For simplicity, let us introduce
\begin{align}
	\omega=\hat{\Omega}_{H}-\hat{\Omega}_{Hc},\qquad \overline{t}=\hat{T}-\hat{T}_{c},\qquad j=J-J_{c},
	\label {4.16}
\end{align}
then eq.(\ref{0722-25}) becomes
\begin{align}
	\omega=c_{10}\overline{t}+c_{20}\overline{t}^2+c_{11}\overline{t}j+c_{30}\overline{t}^3+c_{03}j^3+c_{21}\overline{t}^2j+c_{12}\overline{t}j^2.
	\label {4.17}
\end{align}

According to the equation of state (\ref{4.1})-(\ref{4.3}), one can calculate all coefficients in eq.(\ref{4.17}),
\begin{equation}
	\begin{split}
		c_{10}=-\sqrt{2} \pi,
		\quad c_{20}=\sqrt{6 \left(29 \sqrt{2}-41\right)} \pi ^{3/2} \sqrt{\frac{1}{P}},
		\quad c_{11}=\frac{64}{3} \left(\sqrt{2}-1\right) \pi ^2 P,\\
		\ c_{30}=\frac{9 \left(11 \sqrt{2}-15\right) \pi ^2}{P},
		\quad  c_{03}=-\frac{16384 (\pi P) ^{7/2}}{9 \sqrt{87 \sqrt{2}+123} },\\
		\ c_{21}=-{544 \sqrt{\frac{2P}{5331 \sqrt{2}+7539}} \pi ^{5/2}},
		\quad c_{12}=\frac{2048}{3} \left(4-3 \sqrt{2}\right) \pi ^3 P^2.\label{4.18}
	\end{split}
\end{equation}

The equilibrium condition $\hat{\Omega}_a=\hat{\Omega}_b$ implies $\omega(j_{a},t)=\omega(j_{b},t)$, which in turn yields

\begin{equation}
	\begin{split}
\omega_b-\omega_a&=(j_{b}-j_{a})\big[c_{11}\overline{t}+c_{21}\overline{t}^2+c_{12}\overline{t}(j_{a}+j_{b})+c_{03}(j_{a}^2+j_{a}j_{b}+j_{b}^2)\big]\\
&=0.\label{618-28}
	\end{split}
\end{equation}

\begin{figure}[H]
	\centering
	\includegraphics[width=0.7\textwidth]{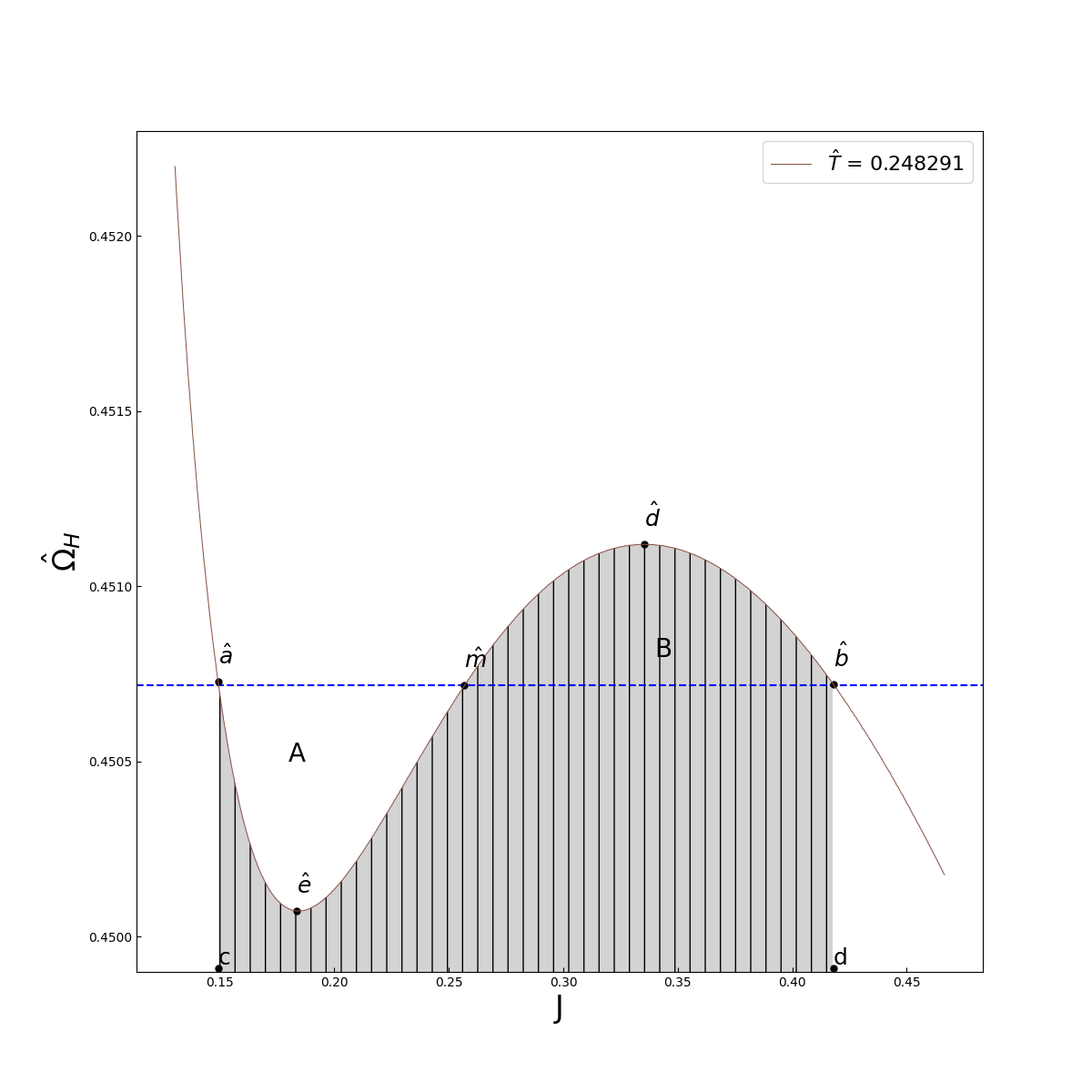}
	\caption{According to the equal area law area($A$)=area($B$), the shaded area should equal to the area of the rectangle $cd\hat{b}\hat{a}$.}
	\label{Fig.4}
\end{figure}

By employing the Maxwell's equal area law, we know the area of the shaded region in Fig.\ref{Fig.4} is equal to the area of the rectangle $cd\hat{b}\hat{a}$ in Fig.\ref{Fig.4}. So we have
\begin{eqnarray}\label{731-30}
\int^{J_b}_{J_a}{\hat\Omega}dJ=(J_b-J_a)\frac{1}{2}(\Omega_b+\Omega_a)
\end{eqnarray}
(The left side of eq.(\ref{731-30}) is the area of the shaded region in Fig. 3 and the right side is the area of the rectangle cdba in Fig.3. The factor $\frac{1}{2}(\Omega_b+\Omega_a)$ comes from the equilibrium condition ${\hat\Omega}_a={\hat\Omega}_b$.)
By using eq. (\ref{4.16}), the eq. (\ref{731-30}) becomes
\begin{eqnarray}
\int^{J_b}_{J_a}{\hat\Omega}dJ
&=&\left[\int^{J_b}_{J_a}({\hat\Omega}-{\hat\Omega}_c)d(J-J_c)\right]+{\hat\Omega}_c[(J_b-J_c)-(J_a-J_c)]\nonumber\\
&=&\left[\int^{j_b}_{j_a}({\hat\Omega}-{\hat\Omega}_c)dj\right]+{\hat\Omega}_c(j_b-j_a)\nonumber\\
&=&\left[\int^{j_b}_{j_a}\omega dj\right]+{\hat\Omega}_c(j_b-j_a)\nonumber\\
&=&(J_b-J_a)\frac{1}{2}({\hat\Omega}_b+{\hat\Omega}_a)\nonumber\\
&=&(j_b-j_a)\frac{1}{2}([{\hat\Omega}_b-{\hat\Omega}_c]+[{\hat\Omega}_a-{\hat\Omega}_c])+(j_b-j_a){\hat\Omega}_c\nonumber\\
&=&(j_b-j_a)\frac{1}{2}(\omega_b+\omega_a)+(j_b-j_a){\hat\Omega}_c.
\end{eqnarray}
Above result implies
\begin{eqnarray}\label{4.19b}
\int^{j_b}_{j_a}\omega dj=(j_b-j_a)\frac{1}{2}(\omega_b+\omega_a).
\end{eqnarray}


 Submitting eq.(\ref{4.17}) into  eq.(\ref{4.19b}), one can obtain
 \begin{equation}
 	\begin{split}
 		&\int_{j_{a}}^{j_{b}}(c_{10}\overline{t}+c_{20}\overline{t}^2+c_{11}\overline{t}j+c_{30}\overline{t}^3+c_{03}j^3+c_{21}\overline{t}^2j+c_{12}\overline{t}j^2) {\rm d}j\\=&c_{10}\overline{t}(j_{b}-j_{a})+c_{20}\overline{t}^2(j_{b}-j_{a})+\frac{1}{2}c_{11}\overline{t}(j_{b}^2-j_{a}^2)+c_{30}\overline{t}^3(j_{b}-j_{a})\\&+\frac{1}{4}c_{03}(j_{b}^4
 -j_{a}^4)+\frac{1}{2}c_{21}\overline{t}^2(j_{b}^2-j_{a}^2)+\frac{1}{3}c_{12}\overline{t}(j_{b}^3-j_{a}^3)\\=&(j_{b}-j_{a})[c_{10}\overline{t}+c_{20}\overline{t}^2+\frac{1}{2}c_{11}\overline{t}(j_{b}+j_{a})
 +c_{30}\overline{t}^3\\&+\frac{1}{4}c_{03}(j_{b}+j_{a})(j_{b}^2+j_{a}^2)+\frac{1}{2}c_{21}\overline{t}^2(j_{b}+j_{a})+\frac{1}{3}c_{12}\overline{t}(j_{b}^2+j_{a}j_{b}+j_{a}^2)]\\=&(j_{b}-j_{a})\frac{1}{2}(\omega_{a}+\omega_{b}),
 		\label{712-2}
 	\end{split}
 \end{equation}

dividing both sides of eq.(\ref{712-2}) by $(j_{b}-j_{a})$ yields
\begin{equation}
	\begin{split}
		&c_{10}\overline{t}+c_{20}\overline{t}^2+\frac{1}{2}c_{11}\overline{t}(j_{b}+j_{a})+c_{30}\overline{t}^3\\&+\frac{1}{4}c_{03}(j_{b}+j_{a})(j_{b}^2+j_{a}^2)+\frac{1}{2}c_{21}\overline{t}^2(j_{b}+j_{a})
+\frac{1}{3}c_{12}\overline{t}(j_{b}^2+j_{a}j_{b}+j_{a}^2)\\=&c_{10}\overline{t}+c_{20}\overline{t}^2+\frac{1}{2}c_{11}\overline{t}(j_{b}+j_{a})+c_{30}\overline{t}^3\\&+\frac{1}{2}c_{03}(j_{b}^3+j_{a}^3)
+\frac{1}{2}c_{21}\overline{t}^2(j_{b}+j_{a})+\frac{1}{2}c_{12}\overline{t}(j_{b}^2+j_{a}^2).
		\label{712-3}
	\end{split}
\end{equation}
Eliminating similar terms on both sides of eq.(\ref{712-3}) gives
\begin{equation}
	\begin{split}
		&\frac{1}{4}c_{03}(j_{b}+j_{a})(j_{b}^2+j_{a}^2)+\frac{1}{3}c_{12}\overline{t}(j_{b}^2+j_{a}j_{b}+j_{a}^2)\\=&\frac{1}{2}c_{03}(j_{b}^3+j_{a}^3)+\frac{1}{2}c_{12}\overline{t}(j_{b}^2+j_{a}^2).
		\label{712-4}
	\end{split}
\end{equation}
Then we can get
\begin{equation}
	\begin{split}
		&\frac{1}{4}c_{03}(j_{b}^3+j_{b}^2j_{a}+j_{b}j_{a}^2+j_{a}^3)+\frac{1}{3}c_{12}\overline{t}(j_{b}^2+j_{a}j_{b}+j_{a}^2)\\=&\frac{1}{2}c_{03}(j_{b}^3+j_{a}^3)+\frac{1}{2}c_{12}\overline{t}(j_{b}^2+j_{a}^2),
		\label{712-5}
	\end{split}
\end{equation}
which implies
\begin{equation}
	\begin{split}
		0=	&\frac{1}{4}c_{03}(j_{a}^3-j_{a}j_{b}^2-j_{a}^2j_{b}+j_{b}^3)+\frac{1}{6}c_{12}\overline{t}(j_{b}^2-2j_{a}j_{b}+j_{a}^2)\\=&\frac{1}{4}c_{03}(j_{b}+j_{a})(j_{b}-j_{a})^2+\frac{1}{6}c_{12}\overline{t}(j_{b}-j_{a})^2.
		\label{712-6}
	\end{split}
\end{equation}

Simplify the above equation, and we find
\begin{equation}
	\begin{split}
		(j_{b}-j_{a})^2\big[\frac{1}{4}c_{03}(j_{a}+j_{b})+\frac{1}{6}c_{12}\overline{t}\big]=0.
		\label{4.20}
	\end{split}
\end{equation}

Denote
$j_{-}\equiv j_{b}-j_{a}=J_{b}-J_{a}, j_{+}\equiv j_{b}+j_{a}$, from eq.(\ref{4.20}), one can solve $j_{+}$ as
\begin{align}
	j_{+}=-\frac{2c_{12}\overline{t}}{3c_{03}},
	\label{618-32}
\end{align}
substituting eq.(\ref{618-32}) into eq.(\ref{618-28}) yields
\begin{align}
	j_{-}=\sqrt{\frac{-4c_{11}\overline{t}+(\frac{4c_{12}^2}{3c_{03}}-4c_{21})\overline{t}^2}{c_{03}}}.\label{4.24}
\end{align}

Near the critical point, the temperature dependence of  the order parameter is
\begin{align}
	\frac{J_{b}-J_{a}}{2}\approx A_{\beta}(\hat{T}-\hat{T}_{c})^{1/2}.
	\label{4.25}
\end{align}

Thus we read $\beta=\frac{1}{2}$.

\subsubsection{Critical exponent of heat capacity}

By definition (\ref{725d3}), the critical exponent of heat capacity can be obtained as follows. Recall that the energy $\hat{M}$ is given by
\begin{align}
	\hat{M}=\frac{\left(a^2+r_{+}^2\right) \left(\frac{r_{+}^2}{l^2}+1\right)}{2 r_{+} \left(1-\frac{a^2}{l^2}\right)^{3/2}},
\end{align}
 the heat capacity $C_{J}$ can be calculated as
\begin{align}
	C_{J}=\big(\frac{\partial\hat{M}}{\partial\hat{T}}\big)_{J}\rvert_{c}=\frac{3}{8P}\neq0.
\end{align}

Due to  the heat capacity neither diverges nor vanishes, it follows that $\alpha$ and $\alpha'$ are both zero, i.e., $\alpha=\alpha'=0$.

\subsubsection{Degree of isothermal compressibility}

The isothermal comprssibility $\kappa{_{T}}$ is defined as
\begin{align}
	\kappa{_{T}}=-\frac{1}{J}\left(\frac{\partial J }{\partial \hat{\Omega}_{H}}\right)_{\hat T},
	\label{4.28}
\end{align}
which diverges at the critical point. Introducing $\widetilde{\omega},\widetilde{t}$ and $\widetilde{j}$ as
\begin{align}
	\widetilde{\omega}=\frac{\hat{\Omega}_{H}-\hat{\Omega}_{Hc}}{\hat{\Omega}_{Hc}},\quad \widetilde{t}=\frac{\hat{T}-\hat{T}_{c}}{\hat{T}_{c}},\quad \widetilde{j}=\frac{J-J_{c}}{J_{c}},
\end{align}
then
\begin{align}
	\hat{\Omega}_{H}=\hat{\Omega}_{Hc}\widetilde{\omega}+\hat{\Omega}_{Hc},\quad \hat{T}=\hat{T}_{c}\widetilde{t}+\hat{T}_{c},\quad J=J_{c}\widetilde{j}+J_{c},\label{4.30}
\end{align}
and
\begin{align}
	\omega=\hat{\Omega}_{Hc}\widetilde{\omega},\quad \overline{t}=\hat{T}_{c}\widetilde{t},\quad j=J_{c}\widetilde{j}.
	\label{4.31}
\end{align}

Eq.(\ref{4.17}) can be rewritten as
\begin{align}
	\widetilde{\omega}=\widetilde{c_{10}}\widetilde{t}+\widetilde{c_{20}}\widetilde{t}^2+\widetilde{c_{11}}\widetilde{t}\widetilde{j}+\widetilde{c_{30}}\widetilde{t}^3+\widetilde{c_{03}}\widetilde{j}^3+\widetilde{c_{21}}\widetilde{t}^2\widetilde{j}+\widetilde{c_{12}}\widetilde{t}\widetilde{j}^2.
	\label{4.32}
\end{align}

Using eq.(\ref{4.28}), we have
\begin{align}
	\frac{-1}{J\kappa{_{T}}}=\frac{1}{\left(\frac{\partial J }{\partial \hat{\Omega}_{H}}\right)_{\hat T}}=\left(\frac{\partial \hat{\Omega}_{H} }{\partial J}\right)_{\hat T}.\label{4.33}
\end{align}

Combining eqs.(\ref{4.25}), (\ref{4.30}), (\ref{4.31}) and (\ref{4.32}) yields
\begin{equation}
	\begin{aligned}
		\frac{-1}{J\kappa{_{T}}}=\left(\frac{\partial \hat{\Omega}_{H} }{\partial J}\right)_{\hat T}=\left(\frac{\partial\hat{\Omega}_{H}}{\partial \widetilde{j}}\frac{\partial \widetilde{j}}{\partial J}\right)_{\hat{T}}=\frac{\hat{\Omega}_{Hc}}{J_{c}}\left(\frac{\partial \widetilde{\omega}}{\partial \widetilde{j}}\right)_{\widetilde{t}}\propto\left(\frac{\partial \omega}{\partial j }\right)_{\overline{t}},
	\end{aligned}
\end{equation}
namely
\begin{align}
	\kappa{_{T}}^{-1}\propto\big(\frac{\partial \omega}{\partial j }\big)_{\overline{t}}= c_{11}\overline{t}+3c_{03}j^2+o(\overline t)= C_{1}\overline{t}+o(\overline t),
\end{align}
where $C_{1}$ is a real constant and  eq.(\ref{4.25}) is used in the last step.
By the definition of $\gamma$ and $\gamma'$ in eq.(\ref{cri4}), we get $\gamma=\gamma'=1$.

\subsection{Free energy and Widom scaling}
\subsubsection{Free energy}
According to the differential expression ${\rm d}G=-S{\rm d}\hat{T}+J{\rm d}\hat{\Omega}_{H}$, at constant angular velocity $\hat{\Omega}_{H}$, we can make use of the Eqs.(\ref{2.3}), (\ref{4.2}) and (\ref{4.3}) to plot $G-\hat{T}$ diagram.
\begin{figure}[H]
	\centering
	\includegraphics[width=0.7\textwidth]{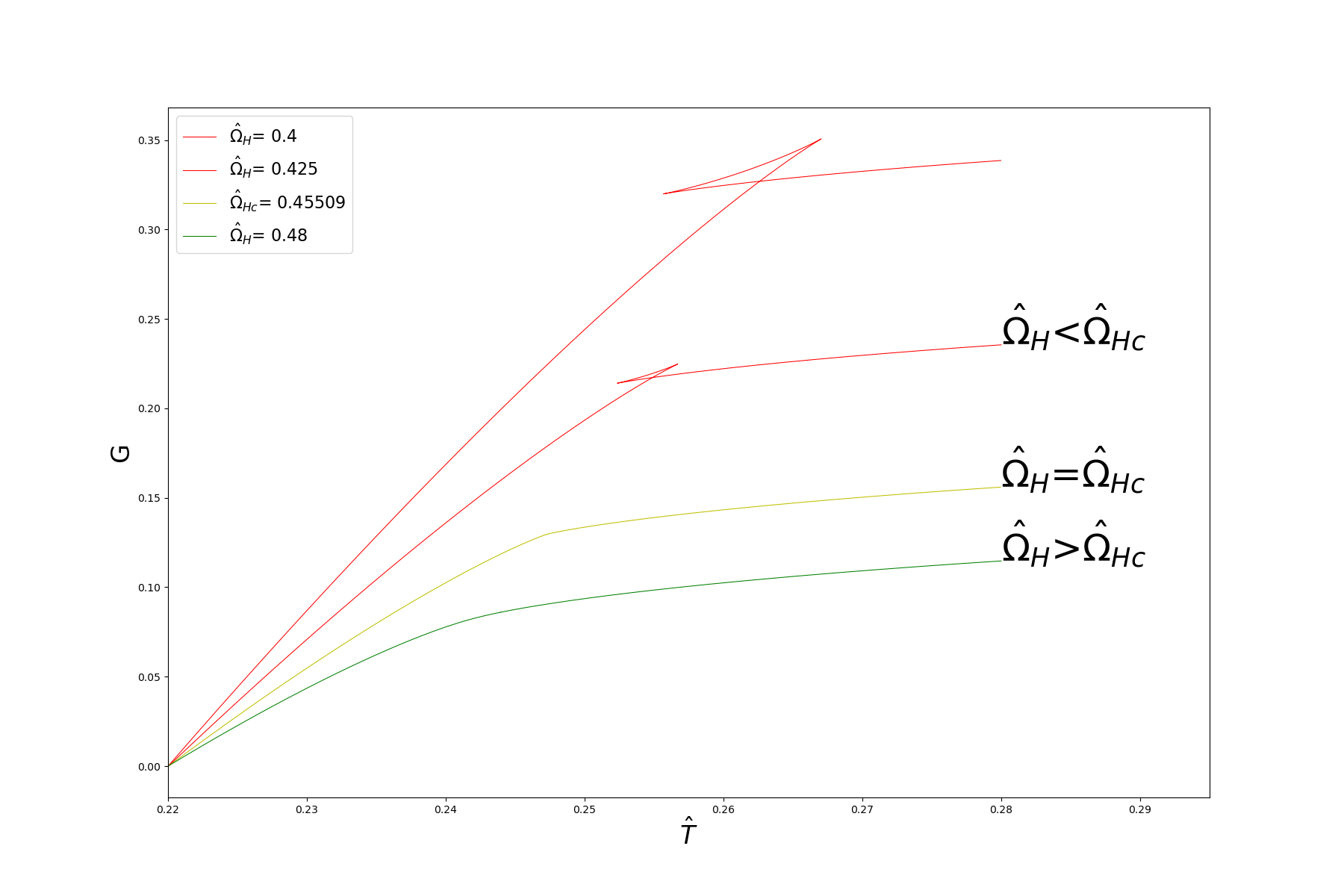}
	\caption{This diagram describes the qualitative behavior of the free energy as a function of temperature for various angular velocity with $l=1$. The angular velocity decreases from bottom to top. The green solid line corresponds to $\hat{\Omega}_{H}>\hat{\Omega}_{Hc}$, the yellow solid line to $\hat{\Omega}_{H}=\hat{\Omega}_{Hc}$, and the remaining red solid lines display $\hat{\Omega}_{H}<\hat{\Omega}_{Hc}$. For $\hat{\Omega}_{H}<\hat{\Omega}_{Hc}$, the existence of swallowtail structure implies there is a first-order phase transition in the system.}
	\label{Fig.5}
\end{figure}
From Fig.\ref{Fig.5}, the swallowtail structure is analogous to the van der Waals system. This further confirms the similarity between the modified thermodynamics and the PVT system.

\subsubsection{Widom scaling law}
A well-known fact is that in the vicinity of the critical point, the free energy can be expressed as a homogenous function, with corresponding homogenous indices $p$ and $q$ \cite{R}. Based on the previous discussions, we see that the critical exponents satisfy the following expected relations:
\begin{align}
	\alpha+2\beta+\gamma=2, \quad \alpha+\beta(\gamma+1)=2,\\
	\gamma(\delta+1)=(2-\alpha)(\delta-1),
	\quad \gamma=\beta(\delta-1).
\end{align}

The free energy has the following scaling symmetry
\begin{align}
	g_{s}(\Lambda^p \epsilon,\Lambda^q j)=\Lambda g_{s}(\epsilon,j),\quad p=\frac{1}{2},\quad q=\frac{3}{4}.
\end{align}

 In terms of $p$ and $q$ \cite{R},  the critical exponets read
\begin{align}
	\alpha=&\frac{2p-1}{q},\\
	\beta=&\frac{1-q}{p},\\
	\gamma=&\frac{2q-1}{p},\\
	\delta=&\frac{q}{1-q}.
\end{align}

The above scaling relations could be regarded as the consistency check for the critical exponents we got in this section.

\section{phase structure on ($P$, $\hat{V}$) section }

\begin{figure}[H]
	\centering
	\includegraphics[width=0.7\textwidth]{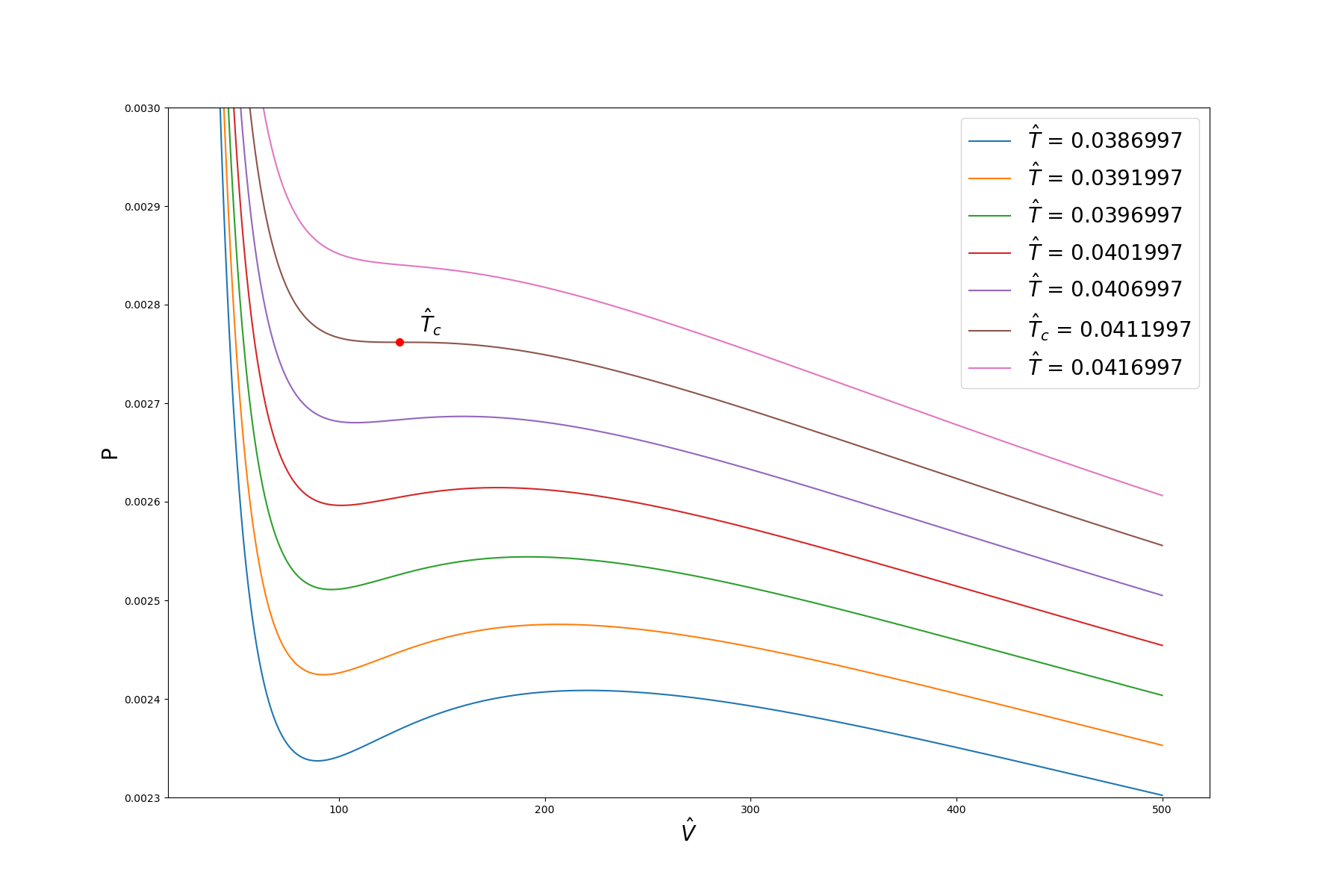}
	\caption{P-$\hat{V}$ diagram of Kerr-AdS black hole with $J=1$. The brown solid line is the critical isotherm and the red dot is the critical point.}
	\label{Fig.6}
\end{figure}

In standard black hole thermodynamics, the phase structure on $(P,V)$ section has been investigated \cite{WL,CWL,ML,GKM,KM}. In this section, we  study the phase structures on ($P $, $\hat{V}$) section in the modified black hole thermodynamics. In Fig.\ref{Fig.6}, we plot the $P-\hat{V}$ diagram corresponding to the Kerr-AdS black hole.

Following the method used in  \cite{KM}, after neglecting all higher order terms of $J$, the equation of state can be written as
\begin{align}
	P=\frac{\sqrt[3]{\frac{\pi }{6}}\hat{T} }{\sqrt[3]{\hat{V}}}-\frac{1}{2\ 6^{2/3} \sqrt[3]{\pi }\hat{V}^{2/3}}+\frac{2 \pi  J^2 \left(4 \sqrt[3]{6} \pi ^{2/3} \hat{T} \sqrt[3]{\hat{V}}+3\right)}{3 \left(\sqrt[3]{6} \pi ^{2/3} \hat{T} \hat{V}^{4/3}+\hat{V}  \right)^2},
	\label{5.1}
\end{align}
where $\hat{V}$ satisfies
\begin{align}
	\hat{V}=\frac{4 \pi  r_{+}^3}{3}+\frac{48 \pi   \left(4 \pi  P r_{+}^2+1\right)J^2}{r_{+} \left(8 \pi  P r_{+}^2+3\right)^2}.
\end{align}
Following the  critical point condition $\frac{\partial P}{\partial \hat V}\big|_{T_C}=0,\frac{\partial^2 P}{\partial \hat V^2}\big|_{T_C}=0$, we can get
\begin{align}
	P_{c}=\frac{0.003}{  J},\quad\hat{V}_{c}=  129.603 J^{3/2},\quad\hat{T}_{c}=\frac{0.041}{  \sqrt{J}}.
\end{align}

Now we consider the phase transition which happens below the critical temperature $\hat T_{c}$. In Fig.\ref{Fig.6}, this van der Waals-like phase structure is clearly visualized. The behavior of the physical variables near its critical point is quantitatively described by the critical exponents. Following a similar method in Section 4, we have
\begin{align}
	\alpha=\alpha'=0, \quad \beta=\frac{1}{2},\quad \gamma=1,\quad \delta=3.
\end{align}
The above results show that the phase structure for the modified black hole thermodynamics in $(P,\hat{V})$ section is almost the same as the standard one \cite{ZMCHR,TB}, the only difference is that all thermal quantities have a deformation factor $\frac{1}{\sqrt{\Xi}}$, which is shown in eq.(\ref{3.1}).
\section{Discussion}

In Kerr-AdS spacetime, there exist two sets of the first law of thermodynamics, because of different choices of observers at infinity. We focus on investigating the critical phenomena of black hole thermodynamics associated with rotating observers. Based on the modified first law and mass formula, we obtain the phase structures in the extended phase space for both the $(\hat{\Omega}_{H},J)$ cross-section and $(P,\hat{V})$ cross-section. In comparison with the previous results \cite{TWY}, the phase structure within the $(\hat{\Omega}_{H},J)$ plane remains analogous to the van der Waals-like phase structure. However, the difference lies in the correspondence of thermodynamic quantities. In previous work \cite{TWY}, the correspondence is $J\rightarrow P$, $\Omega_{H}\rightarrow V$, but in our case it is $\hat{\Omega}_{H}\rightarrow P, J\rightarrow V$. The aforementioned changes result from differences in observers. Compared to non-rotating observers, the energy measured by rotating observers includes rotational kinetic energy part $\sim\hat{\Omega}_{H}J$. This variation is equivalent to performing a Legendre transformation in the extended phase space with respect to the conjugate coordinates $J$ and $\hat{\Omega}_{H}$, which is the reason why the thermodynamic quantities correspond differently. Additionally, we discussed the phase structures on the $(P,\hat{V})$ plane in modified thermodynamics, finding that the obtained results closely resemble those in standard thermodynamics.

\section*{Acknowledgements}

The authors are indebted to Xiaokai He and Naqing Xie for their useful advice. Q. Liu and X. Zhang are supported by the special foundation for Guangxi Ba Gui Scholars and Junwu Scholars of Guangxi University. X.Wu is supported by the National Natural Science Foundation of China Grant No. 12275350. Authors also want to thank the referees for their valuable comments and suggestions. These opinions have greatly enhanced the quality of this paper.

\end{document}